\documentclass[letterpaper, 10 pt, conference]{ieeeconf}  % Comment this line out
\IEEEoverridecommandlockouts                              % This command is only
\usepackage{amsfonts,amsmath,amssymb}
\usepackage{graphicx}

\usepackage{ifthen}
\usepackage{color}
\usepackage{cite}

\usepackage{amsfonts}
\usepackage{amsmath}
\usepackage{amssymb}
\usepackage{color}
\usepackage{graphics}
\usepackage{algorithmicx}
\usepackage{algpseudocode}
\usepackage[boxruled,vlined,linesnumbered,onelanguage]{algorithm2e}
\usepackage{caption}
\usepackage{epsfig}
\usepackage{chngcntr}
\usepackage{dashrule}
\usepackage[position=b]{subcaption}
\usepackage[hyperfootnotes=false]{hyperref}
\usepackage{listings}
\usepackage{color}
\usepackage{authblk}

\usepackage[multiple]{footmisc}

\usepackage{xspace}
\usepackage{tikz}
\usepackage{subcaption}
\usetikzlibrary{calc, positioning, fit, arrows, arrows.meta, shapes.geometric, shapes.symbols, shapes.misc, patterns}

\newcommand{\R}{{\rm I\!R}}

\def\scr#1{{\cal #1}}
\newtheorem{theorem}{Theorem}
\newtheorem{lemma}{Lemma}
\newtheorem{example}{Example}

\newtheorem{corollary}{Corollary}

\newtheorem{remark}{Remark}
\def\qed{ \rule{.08in}{.08in}}

\newtheorem{definition}{Definition}

\def\send#1#2{\stackrel{#1}{\hbox to #2{\rightarrowfill}}}
\def\-{\!\!\!\!\!-}

 \def\qed{ \rule{.1in}{.1in}}

\def\scr#1{{\cal #1}}

\def\qed{ \rule{.1in}{.1in}}

\def\R{{\rm I\!R}} 

\newcounter{seqn}[equation]
\def\theseqn{\arabic{equation}\alph{seqn}}

\def\endseqn{\eqno \@seqnnum
$$\ignorespaces}
\def\@seqnnum{(\theseqn)}

\newskip\mcentering \mcentering=0pt plus 1000pt minus 1000pt

\def\meqalignno#1{
\halign to\displaywidth{
    \hbox to 0pt{\kern\displaywidth\llap{$##$}\hss}\tabskip=\mcentering
    &\hfil$\displaystyle{##}$\tabskip=\mcentering
   &&$\displaystyle{{}##}$\hfil\tabskip=\mcentering
    \crcr
    #1\crcr}}

\def\dspace{\multiply\normalbaselineskip 150
		  \divide\normalbaselineskip 100 \normalbaselines
		  \csname @@normalbaselineskip\endcsname\normalbaselineskip}
\def\sspace{\multiply\normalbaselineskip 200
		 \divide\normalbaselineskip 300 \normalbaselines
		 \csname @@normalbaselineskip\endcsname\normalbaselineskip}
\def\sdspace{\multiply\normalbaselineskip 160
		 \divide\normalbaselineskip 150 \normalbaselines
		 \csname @@normalbaselineskip\endcsname\normalbaselineskip}

\def\@{\tilde}

\def\3dot#1{\buildrel\textstyle...\over#1}

\renewcommand{\qed}{\hfill\blacksquare}

\begin{document}

\author{Yuke Li,\thanks{This work was supported by National Science Foundation grant n.1607101.00 and US Air Force grant n. FA9550-16-1-0290.} and A. Stephen Morse\thanks{Y. Li is with the Department of Political Science, A. Stephen Morse is with the Department of Electrical Engineering, Yale University, New Haven, CT, USA, \{yuke.li, as.morse\}@yale.edu}}

\title{The Power Allocation Game on A Network: A Paradox}

\maketitle

\begin{abstract}
The well-known Braess paradox in congestion games states that adding an additional road to a transportation network may increase the total travel time, and consequently decrease the overall efficiency. Motivated by this, this paper presents a paradox in a similar spirit emerging from another distributed resource allocation game on networks, namely the power allocation game between countries developed in \cite{allocation}. The paradox is that by having additional friends may actually decrease a country's total welfare in equilibrium. Conditions for this paradox to occur as well as some price of anarchy results are also derived. 

%The presentation and analysis of the paradox is made possible by a real-valued utility function representation of countries' preferences which satisfies the two axioms in \cite{allocation}. %This paper also includes certain price of anarchy results for the power allocation game. 


\end{abstract}

\section{Introduction}

In 1969, a paradoxical example was presented in \cite{braess1968paradoxon}, demonstrating that due to selfish behaviors of agents, a measure aimed to increase the efficiency of a transportation network may produce counter-productive effects. Specifically, it was shown that adding a new route to the transportation network can increase the total travel time therein. The concept of price of anarchy \cite{koutsoupias2009worst} was naturally adopted to measure the extent of inefficiency caused by agents' behavior of selfish routing in \cite{lin2004stronger,huang2006efficiency,roughgarden2002bad,youn2008price,roughgarden2005selfish,lin2011stronger}. For example, \cite{roughgarden2002bad} obtains lower and upper bounds for the price of anarchy in the congestion game on any transportation network. An optimal network design problem was then formulated and extensively studied in \cite{chen1991network,friesz1992simulated,leblanc1975algorithm,roughgarden2001designing,yang1998models,magnanti1984network,brown2017studies, poorzahedy1982approximate, suwansirikul1987equilibrium}, where the motivation behind this problem was the potential interest of policy makers in designing a transportation network with the goal of minimizing the price of anarchy involved.

%In the following work such as \cite{valiant2006braess, huang2006efficiency, braess2005paradox, dafermos1984some}, conditions regarding agents' preferences are derived for Braess paradox to arise in the congestion game on a general transportation network. In \cite{kameda2000braess}, similar paradoxes can also exist in other networks such as distributed computer systems. 
 
This paper proposes a similar paradox that arises in another distributed, resource allocation game on networks, where countries ``allocate'' their power among their friends and adversaries, namely the power allocation game (PAG). This is a distributed, resource allocation game on a signed graph.\cite{balancing}

This paper focuses on the analysis of the paradox by examining the case of a loss in welfare countries may suffer due to certain changes in the networked environment that was supposed to increase its utility. For example, having additional friends in the environment may prevent a country from achieving its optimal welfare.  
 
Obviously, an utility function is needed to the definition of countries' ``welfare'' from power allocation and to the introduction of the paradox; a certain family of utility functions that satisfy the two axioms that model countries' preferences for the ``power allocation matrices'' in \cite{allocation} and \cite{survival} will be introduced and used throughout the paper.

The paper is structured as the following. In Section II, the set up of the PAG is reviewed, along with a discussion of a family of utility functions to model countries' preferences for ``power allocation matrices'', and a paradox is identified. In Section III, conditions for this paradox to occur are derived. The paper concludes with a discussion of the upper and lower bounds for the price of anarchy in the power allocation game in a general networked environment.

\section{The PAG and the Paradox}

\subsection{Basic Idea} By the  {\em power allocation game} or PAG is meant a distributed resource allocation game between $n$ countries with labels in $\mathbf{n}= \{1,2,\ldots,n\}$\cite{allocation}. The game is formulated on a simple, undirected,  signed graph $\mathbb{G}$ called ``an environment graph'' \cite{survival} whose $n$ vertices correspond to the countries and whose $m$ edges represent relationships between countries. An edge between distinct vertices $i$ and $j$, denoted by $(i,j)$, is labeled with a plus sign if countries $i$ an $j$ are friends and with a minus sign if countries $i$ and $j$ are adversaries.
For each $i\in\mathbf{n}$, $\scr{F}_i$ and $\scr{A}_i$ denote the sets of labels of country $i$'s friends and adversaries respectively; it is assumed that $i\in\scr{F}_i$ and that $\scr{F}_i$ and $\scr{A}_i$ are disjoint sets.
Each country $i$ possesses a nonnegative quantity $p_i$ called the {\em total power} of country $i$. An allocation of this power or {\em strategy} is a nonnegative $n\times 1$ row vector $u_i$ whose $j$ component $u_{ij}$ is that part of $p_i$ which country $i$ allocates under the strategy to either support country $j$ if $j\in\scr{F}_i$ or to demise country $j$ if $j\in\scr{A}_i$; accordingly $u_{ij}= 0$ if $j\not\in\scr{F}_i\cup\scr{A}_i$. The goal of the game is for each country to choose a strategy which contributes to the demise of all of its adversaries and to the support of all of its friends.

Each set of country strategies $\{u_i,\;i\in\mathbf{n}\}$ determines an
 $n\times n$ matrix  $U$ whose $i$th row is $u_i$. Thus $U = [u_{ij}]_{n\times n}$ is a nonnegative matrix such that, for each $i\in\mathbf{n}$, $u_{i1}+u_{i2}+\cdots +u_{in} = p_i$. Any such matrix is called a {\em strategy matrix}  and $\scr{U}$ is the set of all $n\times n$ strategy matrices.
 
\subsection{Multi-front Pursuit of Survival}

In \cite{allocation} and \cite{survival}, how countries allocate their power in the support of \emph{the survival} of its friends and the demise of that of its adversaries is studied, which is in line with the fundamental assumptions about countries' behavior in classical international relations theory.\cite{waltz1979theory} These facts are accounted for by the following additional formulations: 

Each strategy matrix $U$ determines for each $i\in\mathbf{n}$, the {\em total support} $\sigma_i(U)$ of
 country $i$ and the {\em total threat}  $\tau_i(U)$ against country $i$.  Here
 $\sigma_i:\scr{U}\rightarrow\R$ and $\tau_i:\scr{U}\rightarrow \R$ are non-negative
 valued maps defined by
$U\longmapsto \sum_{j\in\scr{F}_i}u_{ji} +\sum_{j\in\scr{A}_i}u_{ij}$
 and $U\longmapsto \sum_{j\in\scr{A}_i}u_{ji}$
  respectively. Thus country $i$'s total support is the sum of the amounts of power
   each of country $i$'s friends
  allocate to its support plus the sum of the amounts of power country $i$ allocates  to the
  destruction of  all of its adversaries. Country $i$'s total threat, on the other hand, is the sum of the amounts of power
  country $i$'s adversaries allocate to its demise. These allocations in turn determine
   country $i$'s {\em state} $x_i(U)$ which may be safe, precarious, or unsafe depending on the relative
    values of $\sigma _i(U)$ and $\tau_i(U)$. In particular, %$x_i:\scr{U}\rightarrow $
    %\{safe, precarious, unsafe\} is the map defined so that
     $x_i(U) = $ safe
    if $\sigma_i(U)>\tau_i(U)$, $x_i(U)=$ precarious if $\sigma_i(U)=\tau_i(U)$, or
    $x_i(U) = $ unsafe if $\sigma_i(U)<\tau_i(U)$.

In playing the PAG, countries select individual strategies in accordance with certain weak and/or strong preferences.
A sufficient set of conditions for  country $i$ to {\em weakly prefer} strategy
matrix $V\in\scr{U}$ over strategy matrix $U\in\scr{U}$ are as follows
\begin{enumerate}
\item  For all $j\in\scr{F}_i$
 either $x_j(V)\in$ \{safe, precarious\}, or $x_j(U) = $ unsafe, or both.
 \item For all $j\in\scr{A}_i$ either $x_j(V) \in $ \{unsafe, precarious\}, or $x_j(U) = $ safe, or both.
\end{enumerate}
Weak preference  by country $i$ of  $V$ over $U$ is denoted by $U \preceq V$.

Meanwhile, a sufficient condition for country
 $i$ to be {\em indifferent} to the choice between $V$ and $U$ is  that $x_i(U)=x_j(V)$ for all $j\in\scr{F}_i\cup\scr{A}_i$.  This is denoted by $V\sim U$.

Finally, a sufficient condition for country $i$  to {\em strongly prefer} $V$ over $U$ is that $x_i(V)$ be a  safe or precarious  state and $x_i(U)$ be an unsafe state. Strong preference by country $i$ of $V$ over $U$ is denoted by $U \prec V$.

\subsection{Preferences: Utility function representation}

A country $i$ will derive a certain amount of utility or welfare from a strategy profile of the PAG assuming a certain networked environment; let $i$'s utility be a function $f_i: \mathcal{U} \longrightarrow \mathbb{R}$. A family of utility functions with the following three properties satisfies the two preference axioms \emph{and} makes possible a total order of the power allocation matrices in $\mathcal{U}$. For more details about how this is achieved, please refer to \cite{networkgame}. 

\begin{enumerate}
\item Country $i$ receives a two-valued pairwise utility from each of its friends and adversaries, $t_{ij}(x_{j}(U))$. The specific values of the pairwise utilities can be regarded as proxies of the relative importances attached to each friend and adversary relation. 

Pairwise utility $t_{ij}: \{\text{safe}, \text{precarious}, \text{unsafe}\} \longrightarrow \{t_{ij}(1), t_{ij}(0)\}$ where $t_{ij}(1) \geq t_{ij}(0)$. $t_{ij}(1), t_{ij}(0) \in \mathbb{R}$ is defined as follows. 

If $j \in \mathcal{F}_{i}$, $t_{ij}(1)$ and $t_{ij}(0)$ respectively stands for $i$'s pairwise utility from $j$ when $\sigma_{j}(U) \geq \tau_{j}(U)$ and $\sigma_{j}(U) < \tau_{j}(U)$;

If $j \in \mathcal{A}_{i}$, $t_{ij}(1)$ and $t_{ij}(0)$ respectively stands for $i$'s pairwise utility from $j$ when $\sigma_{j}(U) \leq \tau_{j}(U)$ and $\sigma_{j}(U) > \tau_{j}(U)$.

\item Country $i$'s total utility $f_{i}(U)$ will only be equal to $t_{ii}^0$ if it has not survived itself.

\item Once country $i$ has survived, its total utility $f_{i}(U) > t_{ii}^0$, and will be a nondecreasing function of $t_{ij}(x_{j}(U))$ for any $j \in \mathcal{F}_{i} \cup \mathcal{A}_{i}$.

\end{enumerate}

For simplicity, this paper's focus is on a subset of utility functions in this family; for any utility function in this subset, its maximum is attained only when all its friends are safe/precarious, and all its adversaries are unsafe/precarious. 

After country $i$'s self-survival threshold is fulfilled, the function $f_{i}(U)$ exhibits a jump discontinuity. Let the set of $i$'s friends who are safe/precarious under $U$ be $\mathcal{F}^{1}_{i}(U)$ and the set of $i$'s adversaries who are unsafe/precarious be $\mathcal{A}^{1}_{i}(U)$. An example is $$ f_{i}(U) = \\\begin{cases}
t_{ii}(0) & x_{i}(U) = \text{unsafe} \\
\sum_{i \in \mathcal{F}^{1}_{i}(U) \cup \mathcal{A}^{1}_{i}(U)}t_{ij}(1)  & x_{i}(U) \in \{\text{safe, precarious}\}\\
%0 & j \not\in \{i\} \cup \mathcal{F}_{i} \cup \mathcal{A}_{i}
 \end{cases}$$

Some functions in this family may exhibit more jump discontinuities; for instance, countries may additionally prioritize the survival of some friends. An utility function where country $i$ prioritizes the survival of all its friends before demising the survival of all its adversaries is below: $$ f_{i}(U) = \begin{cases}  
t_{ii}(0) & x_{i}(U) = \text{unsafe}\\
\sum_{j \in \mathcal{F}^{1}_{i}(U)} t_{ij}(1)  & x_{i}(U) \in \{\text{safe, precarious}\}\\ 
\sum_{i \in \mathcal{F}^{1}_{i}(U) \cup \mathcal{A}^{1}_{i}(U)}t_{ij}(1) & x_{i}(U) \in \{\text{safe, precarious}\},\\ & \forall j \in \mathcal{F}_i 
\end{cases}$$

%Therefore, this discussion of utility functions will naturally precede the analysis of both the paradox and the price of anarchy in the power allocation game. Another implication that can be of independent interest is that the utility function representation of countries' preferences thus extends the partial order of all power allocation matrices defined by the two axioms into a total order. 

\subsection{Definition: A Paradox}

Let country $i$'s \emph{optimal welfare from power allocation} be the maximum utility $i$ can derive from a power allocation matrix $U \in \mathcal{U}$. Denote it as $$f^{*}_{i}(U)$$ 

Since country $i$'s optimal welfare from power allocation must differ by environments, we define a \emph{pairwise utility} $i$ receives from having every other country $j$ (other than $i$) for each of the following four conditions:

\begin{enumerate}
\item $t_{ij}^{\mathcal{F}}(1)$: $j \in \mathcal{F}_{i}$ and $x_{j}(U) \in \{\text{safe}, \text{precarious}\}$.
\item $t_{ij}^{\mathcal{F}}(0)$: $j \in \mathcal{F}_{i}$ and $x_{j}(U) = \text{unsafe}$.
\item $t_{ij}^{\mathcal{A}}(1)$: $j \in \mathcal{A}_{i}$ and $x_{j}(U) \in \{\text{unsafe}, \text{precarious}\}$.
\item $t_{ij}^{\mathcal{A}}(0)$: $j \in \mathcal{A}_{i}$ and $x_{j}(U) = \text{safe}$.
\end{enumerate}

Let the two environments be with the same set of countries and one difference between the environments is that for country $i$, $\mathcal{F}_{i} \subset \mathcal{\overline{F}}_{i}$. A \emph{paradox} is said to occur if given two PAGs  $\Gamma$ and $\overline{\Gamma}$, country $i$ can obtain a higher optimal welfare from the former where it has fewer friends.

% \begin{figure}
% \begin{subfigure}{0.2\textwidth}
% \begin{tikzpicture}[x=9em,y=-8em][scale=0.3]
% 	\drawnodex{-1,-1/2}{v1}{left}{v1}{above}{$8$}
% 	\drawnodex{-1,1/2}{v2}{left}{v2}{below}{$6$}
% 	\drawnodex{0,0}{v3}{below}{v3}{above}{$1$}

%     \drawgreenedge{v1}{v3}%{$w_{1,3}$}{$w_{3,1}$}
% 	\drawgreenedge{v2}{v3}%{$w_{2,3}$}{$w_{3,2}$}
% 	\drawrededge{v1}{v2}%{$w_{1,2}$}{$w_{2,1}$}
% \end{tikzpicture}
% \caption{Environment I}
% \end{subfigure}
% \begin{subfigure}{0.2\textwidth}
% \begin{tikzpicture}[x=9em,y=-8em][scale=0.3]
% \drawnodex{-1,-1/2}{v1}{left}{v1}{above}{$8$}
% 	\drawnodex{-1,1/2}{v2}{left}{v2}{below}{$6$}
% 	\drawnodex{0,0}{v3}{below}{v3}{above}{$1$}
%     \drawgreenedge{v1}{v3}%{$w_{1,3}$}{$w_{3,1}$}
% 	\drawrededge{v2}{v3}%{$w_{2,3}$}{$w_{3,2}$}
% 	\drawrededge{v1}{v2}%{$w_{1,2}$}{$w_{2,1}$}
% \end{tikzpicture}
% \caption{Environment II}
% \end{subfigure}
% \caption{A Paradox for country 3}
% \end{figure}

\begin{example} The following will illustrate an example of a paradox, which further motivates the main results in Section III.
The parameters of the first environment in Figure 1(a) is: \begin{enumerate}
\item Set of countries: $\mathbf{n} = \{1, 2, 3\}$. 
\item Their power: $p = [8\quad 6 \quad1]$.
\item Their relations are: $r(1,2) = \text{adversary}$, and $r(2,3) = r(1,3) = \text{friend}$.
\end{enumerate} The parameters of the second environment in Figure 1(b) is:\begin{enumerate}
\item Set of countries: $\mathbf{n} = \{1, 2, 3\}$. 
\item Their power: $p = [8 \quad 6 \quad 1]$.
\item Their relations are: $r(1,2) = r(2,3) = \text{adversary}$, and $r(1,3) = \text{friend}$.
\end{enumerate}

In the first environment, country 3 has a friend relation with both country 1 and country 2, which are adversaries. Any pure strategy Nash equilibrium from the PAG in this environment will predict country 1 to be safe, country 2 to be unsafe/precarious, and country 3 to be safe. If country 3's utility from having country 2 as a safe/precarious friend is lower than having it as an unsafe/precarious adversary $$t_{32}^{\mathcal{F}}(1) < t_{32}^{\mathcal{A}}(1),$$ it will prefer the second environment where it turns against country 2 with country 1 and gains a higher optimal welfare (with country 2 being unsafe/precarious and countries 1 and 3 being safe in any equilibrium). By assumption, $t_{32}^{\mathcal{F}}(1) \leq t_{32}^{\mathcal{F}}(1)$. Therefore, country 3's optimal welfare in the first environment is always lower than that in the second environment, which constitutes the paradox. 

\end{example}

\section{Main Results}

%The argument for the roles of friends in a country's survival proceeds with  Theorem 1, which shows that by having additional friends in an environment will not jeopardize a country's survival in equilibrium.  

This section presents results on the conditions for the stated paradox to occur. In particular, the conditions involve a comparison of the roles of friends in a country's survival and, after the self-survival is fulfilled, in its attainment of its optimal welfare. 

\begin{theorem}

If a country can survive in the equilibria of the PAG assuming a certain networked international environment, then it can survive in the equilibria of another PAG assuming an environment with additional friends than before, but not vice versa. 

Let two PAGs be $\Gamma$ and $\overline{\Gamma}$, where the only difference between the two environments is that for a country $i$ $\mathcal{F}_i \subset \mathcal{\overline{F}}_{i}$. If $\forall U^* \in \mathcal{U}^*$,  $x_i(U^*) \in \{\text{safe}, \text{precarious}\}$, then it must be that $\forall \overline{U}^* \in \mathcal{\overline{U}}^*$, $x_i(\overline{U}^*) \in \{\text{safe}, \text{precarious}\}$. 

\end{theorem}

\noindent {\it Proof of Theorem 1:} In $\Gamma$, if for a country $i$, $\forall U^* \in \mathcal{U}^*$, $x_i(U^*) \in \{\text{safe}, \text{precarious}\}$, by definition $$\sum_{j \in \mathcal{F}_i}u^*_{ji} + \sum_{j \in \mathcal{A}_i}u^*_{ij} \geq \sum_{j \in \mathcal{A}_i}u^*_{ji}.$$ 

%There also must hold that $$ \text{inf} \{\sigma_{i}(U^*): U^* \in \mathcal{U}^*\} \leq \text{inf} \{\sigma_{i}(\overline{U}^*): U^* \in \mathcal{\overline{U}}^*\}.$$ 

%and 

%and that $$ \text{inf} \{\sigma_{i}(U^*): \overline{U}^* \in \mathcal{\overline{U}}^*\} \geq \text{inf} \{\sigma_{i}(\overline{U}^*): U^* \in \mathcal{U}^*\}.$$

Given that $\mathcal{F}_i \subset \mathcal{\overline{F}}_{i}$, any equilibrium allocations in the two games satisfy the following: $$\sum_{j \in \mathcal{F}_i} \overline{u}^*_{ji} + \sum_{j \in \mathcal{A}_i}\overline{u}^*_{ij} \geq \sum_{j \in \mathcal{F}_i} u^*_{ji} + \sum_{j \in \mathcal{A}_i}u^*_{ij}$$ and $$\sum_{j \in \mathcal{F}_i} u^*_{ji} + \sum_{j \in \mathcal{A}_i}u^*_{ij} \geq \text{sup} \{\tau_{i}(U^*): U^* \in \mathcal{U}^*\}.$$ 
Note that $$\text{sup} \{\tau_{i}(U^*): U^* \in \mathcal{U}^*\} = \text{sup} \{\tau_{i}(\overline{U}^*): \overline{U}^* \in \mathcal{\overline{U}}^*\},$$ because the only difference between the two environments lies in the number of $i$'s friends. 

Therefore, in any equilibrium $\overline{U}^* \in \mathcal{\overline{U}^*}$ of $\overline{\Gamma}$, it must also be that $$\sum_{j \in \mathcal{F}_i}\overline{u}^*_{ji} + \sum_{j \in \mathcal{A}_i}\overline{u}^*_{ij} \geq \sum_{j \in \mathcal{A}_i}\overline{u}^*_{ji}.$$ 

In other words, $i$ must also survive in any equilibrium in the game $\overline{\Gamma}$ with more friends in the new environment. 

$\qed$

\begin{remark}
The existence of multiple equilibria in the PAG is the reason in Theorem 1 for the comparison between two games, in all of whose equilibria country $i$ survives. Intuitively, the reverse of Theorem 1 may not be true, with the logic being that country $i$ may not gather the level of support in the previous environment to survive in a new environment with some of its former friends becoming nonexistent. Theorem 1 also holds in the case where country $i$ has some of the former adversaries turn nonexistent or new friends, which means when $\mathcal{F}_i \subset \mathcal{\overline{F}}_{i}$ and $\mathcal{\overline{A}}_i \subset \mathcal{A}_{i}$ hold.\end{remark} %The same proof can be used for this extension. 

Next a necessary condition and a sufficient condition will respectively be provided that a country may achieve a lower optimal welfare in the equilibria of the power allocation game in a new networked environment with more friends than in the previous environment. 

%This paradox happens as a country moves beyond the pursuit of survival to the pursuit of success.  

% To be precise with regard to ``having additional friends'', it may mean that with older friends unchanged, a country may now have new friends with countries previously it has no relations or was adversarial with or a combination of both. 

% In addition, to be precise with regard to ``the decrease in utility'', Theorem 3 states a necessary condition about the previous environment, characterized in terms of countries' power configuration and utility function features (preference structures) for a version of the paradox to occur, that is, having additional friends can lower the \emph{maximum utility} a country may receive in the power allocation game. In other words, it may limit a country from attaining success. 

\begin{theorem}[Necessary Condition]

A necessary condition for the above stated paradox to occur is that there exists at least a country which derives a higher utility from having another country as an unsafe/precarious adversary than as a unsafe friend. 

Given two games $\Gamma$ and $\overline{\Gamma}$, the only difference between the two underlying environments is $\mathcal{F}_i \subset \mathcal{\overline{F}}_{i}$. If there exists country $i \in \mathbf{n}$, $f^{*}_{i}(U) > \overline{f}^{*}_{i}(\overline{U})$, then there must exist country $j \neq i$ such that $t_{ij}^{\mathcal{F}}(0) < t_{ij}^{\mathcal{A}}(1)$. 

\end{theorem}

\noindent {\it Proof of Theorem 2:} Suppose to the contrary. That is to say, for any country $i$, $t_{ij}^{\mathcal{F}}(0) \geq t_{ij}^{\mathcal{A}}(1)$, $j \neq i$, which means that for any country $i$ the utility of having any other country as an unsafe friend exceeds that of having it as an unsafe/precarious adversary. 

Given a random environment, let the optimal welfare country $i$ can obtain from the PAG $\Gamma$ assuming this environment be $f^*_{i}(U)$. 

Let an alternative environment be $\overline{\Gamma}$ be such that the only difference from $\Gamma$ is that $\mathcal{F}_i \subset \mathcal{\overline{F}}_{i}$. Let the optimal welfare country $i$ can obtain from the PAG $\Gamma$ assuming this environment be $\overline{f}^*_{i}(U)$

Then there must hold $$f^{*}_{i}(U) > \overline{f}^{*}_{i}(\overline{U})$$ because for any of $i$'s new friends $j$, even when $x_{j}(\overline{U}) \in \{\text{unsafe},\text{precarious}\}$, $t_{ij}^{\mathcal{F}}(0) \geq t_{ij}^{\mathcal{A}}(1)$.  

Then $i$'s having more friends will not decrease its optimal welfare from power allocation. 
 
Therefore, in order to the stated paradox to occur, there must exist another country $j \neq i$ such that $t_{ij}^{\mathcal{F}}(0) < t_{ij}^{\mathcal{A}}(1)$.  

$\qed$

%$t_{ij}^{\mathcal{F}}(0) \geq t_{ij}^{\mathcal{A}}(0)$, $j \in \mathbf{n} - \{i\}$.

%In games $\overline{\Gamma}$ and $\Gamma$, we know that $f^{*}_{i}(U) \leq \overline{f}^{*}_{i}(\overline{U})$. 

%Let equilibrium strategies of country $i$ and its friends and adversaries in $\Gamma$ assuming the previous environment remain the same in $\overline{\Gamma}$ assuming the new environment. 

% Another version of the paradox states more generally that by having additional friends in the environment, a country may find itself in a new equilibrium with fewer utility than in a previous equilibrium. Theorem 3 states a sufficient condition about the previous environment for this paradox to occur. A difference from Theorem 3 is that here having additional friends can mean turning former adversaries into friends. 

\begin{theorem}[Sufficient Condition]

A sufficient condition for the stated paradox to occur is that there exists at least a country which derives a higher utility from having another as an unsafe/precarious adversary than as a safe/precarious friend and the total power of these two countries is smaller than that of all other countries in the environment. 

For country $i \in \mathbf{n}$, suppose that there exists another country $j \neq i$ such that $t_{ij}^{\mathcal{A}}(1) > t_{ij}^{\mathcal{F}}(1)$, and that $p_i + p_j \leq \displaystyle\sum_{k \in \mathbf{n} - \{i,j\}}p_i$. Then there can be constructed two different environments in which the PAG takes place $\Gamma$ and $\overline{\Gamma}$ where the only difference between the environments is that for a country $i$ $\mathcal{F}_i \subset \mathcal{\overline{F}}_{i}$. The stated paradox will then occur for country $i$ as it switches from $\Gamma$ to $\overline{\Gamma}$, which means that $f^*_i(U^*) < \overline{f}^*_i(\overline{U}^{*})$.

\end{theorem}

\noindent {\it Proof of Theorem 3:} Let the environment of the PAG $\Gamma$ be such that all countries other than $i$ are adversaries with $j$. $i$ is a friend with all of the other countries including $j$. And let the environment of the PAG $\overline{\Gamma}$ be such that where all countries are adversaries with $j$, and $i$ is a friend with all of the other countries excluding $j$.

Since $$p_i + p_j \leq \displaystyle\sum_{k \in \mathbf{n} - \{i,j\}}p_i,$$ there must hold that in any equilibrium $U^* \in \mathcal{U}^*$ of $\Gamma$, $$x_{j}(U^*) \in \{\text{unsafe, precarious}\},$$ and $$\forall k \neq j, x_{k}(U^*) \in \{\text{safe}, \text{precarious}\}.$$ The same must hold for in any equilibrium $\overline{U}^* \in \mathcal{\overline{U}}^*$ of $\overline{\Gamma}$.

To country $i$, country $j$ is an unsafe/precarious friend in $\Gamma$ and an unsafe/precarious adversary in $\overline{\Gamma}$. 

Since $t_{ij}^{\mathcal{A}}(1) > t_{ij}^{\mathcal{F}}(1) \geq t_{ij}^{\mathcal{F}}(0)$, it must be that $f^*_i(U^*) < \overline{f}^*_i(\overline{U}^*)$, with all other pairwise utilities from other neighbors being equal.  

Therefore, having additional friends than before will decrease country $i$'s optimal welfare from power allocation. 

$\qed$

Theorem 3 extends to the case where a country may have a subset of countries in the environment, each of which satisfies the stated condition. Thus Corollary 1 immediately follows. Then when the conditions in Corollary holds, having more friends from this subset will only decrease its optimal welfare from power allocation for a country.

\begin{corollary}

If there exists at least a country which derives a higher utility from having any other in a subset of countries (which it is not a member of) as an unsafe/precarious foe than as a safe/precarious friend and the total power of this country and those in the subset is smaller than that of all other countries in the environment, the stated paradox will occur. 

For country $i \in \mathbf{n}$, suppose that there exists a subset of countries $\mathcal{S}$ such that $i \not\in \mathcal{S}$ and for any $j \in \mathcal{S}$, $t_{ij}^{\mathcal{A}}(1) > t_{ij}^{\mathcal{F}}(1)$, and that $p_{i} + \displaystyle\sum_{j \in \mathcal{S}} p_j \leq \displaystyle\sum_{k \in \mathbf{n} -\{i\} - \mathcal{S}}p_k$. Then there exists two games assuming different environments $\Gamma$ and $\overline{\Gamma}$ where the only difference between the environments is that for a country $i$ $\mathcal{F}_i \subset \mathcal{\overline{F}}_{i}$, and where an $U^{*} \in \mathcal{U}^{*}$ and an $\overline{U}^{*} \in \mathcal{\overline{U}}^{*}$ exist such that $f^*_i(U^*) < \overline{f}^*_i(\overline{U}^{*})$.

\end{corollary}

\noindent {\it Proof of Corollary 1:} Let the environment of the PAG $\Gamma$ be such that all countries other than $i$ are adversaries with any country $j$ in $\mathcal{S}$. $i$ is a friend with all of the other countries including countries in $\mathcal{S}$. Let the environment of the PAG $\overline{\Gamma}$ be such that all countries are adversaries with countries in $\mathcal{S}$, and $i$ is a friend with all of the other countries excluding those in $\mathcal{S}$.

Since $$p_{i} + \displaystyle\sum_{j \in \mathcal{S}} p_j \leq \displaystyle\sum_{k \in \mathbf{n} -\{i,\mathcal{S}\}}p_k,$$ there must hold that in any equilibrium $U^* \in \mathcal{U}^*$ of $\Gamma$, $$\forall j \in \mathcal{S}, x_{j} = \{\text{unsafe,precarious}\},$$ and $$\forall k \not\in \mathcal{S}, x_{k} \in \{\text{safe}, \text{precarious}\}.$$ The same must also hold in any equilibrium $\overline{U}^* \in \mathcal{\overline{U}}^*$ of $\overline{\Gamma}$. 

To country $i$, any country $j \in \mathcal{S}$ is an unsafe/precarious friend in $\Gamma$ and an unsafe/precarious adversary in $\overline{\Gamma}$.

Since $\forall j \in \mathcal{S}$, $t_{ij}^{\mathcal{A}}(1) > t_{ij}^{\mathcal{F}}(1) \geq t_{ij}^{\mathcal{F}}(0)$, there holds that $f^*_i(U^*) < \overline{f}^*_i(\overline{U}^*)$. Therefore, having additional friends than before will decrease country $i$'s optimal welfare from power allocation.

$\qed$

% \begin{remark}

% When countries pursue goals beyond self-survival such as success, it may be even be to their advantages to have a few adversaries, depending on their preference structure and, especially, the power configuration in the environment.  This also suggests that certain existing alignments may lock in commitments and force some countries to choose sides between friends. 

% \end{remark}

\section{The price of anarchy results}

In this section we compare the implications of different networked international environments for the total welfare of countries in the power allocation game, and the commonly defined price of anarchy concept will be used for the analysis. 

\begin{definition}[Price of Anarchy Concept\cite{koutsoupias2009worst}]
$$\frac{\displaystyle\max_{U\in \mathcal{U}} \displaystyle\sum_{i \in \mathbf{n}}f_{i}(U)}{\displaystyle\min_{U \in \mathcal{U^*}}\displaystyle\sum_{i \in \mathbf{n}}f_{i}(U^*)}$$

\end{definition}

\begin{lemma}

In any PAG $\Gamma$, at least a country survives in any $U \in \mathcal{U}$. Note that this holds regardless of whether $\mathcal{U}$ is an equilibrium. 

In $\Gamma$, $\forall U \in \mathcal{U}$, $\exists i \in \mathbf{n}$ such that $x_i(U) \in \{\text{safe}, \text{precarious}\}$. 

\end{lemma}

\noindent {\it Proof of Lemma 1:} The proof is by contradiction. Given an $U \in \mathcal{U}$, suppose that $\forall i \in \mathbf{n}$, $x_i(U) = \text{unsafe}$. That is to say that,$$\forall i \in \mathbf{n}, \sum_{j \in \mathcal{F}_{i}}u_{ji} + \sum_{j \in \mathcal{A}_{i}}u_{ij} < \sum_{j \in \mathcal{A}_{i}}u_{ji}.$$ Equivalently, $$\forall i \in \mathbf{n}, p_{i} - \sum_{j \in \mathcal{F}_{i}}u_{ji} + \sum_{j \in \mathcal{F}_{i}}u_{ji} < \sum_{j \in \mathcal{A}_{i}}u_{ji}.$$

Summing from 1 to $n$ in $\mathbf{n}$ gives the following, $$\sum_{i \in \mathbf{n}}p_i - \sum_{i \in \mathbf{n}}(\sum_{j \in \mathcal{F}_{i}}u_{ji} - \sum_{j \in \mathcal{F}_{i}}u_{ij}) <  \sum_{i \in \mathbf{n}}\sum_{j \in \mathcal{A}_{i}}u_{ji}.$$ Note that, $$\sum_{i \in \mathbf{n}}(\sum_{j \in \mathcal{F}_{i}}u_{ji} - \sum_{j \in \mathcal{F}_{i}}u_{ij}) = \sum_{\{i, j\} \in \mathcal{R}_{\mathcal{F}}}(u_{ji} -  u_{ij}) + (u_{ij} -  u_{ji}) = 0.$$

Then there holds that $$\sum_{i \in \mathbf{n}}p_i < \sum_{i \in \mathbf{n}}\sum_{j \in \mathcal{A}_{i}}u_{ji}.$$

However, by each country's total power constraint, it must be the case that $$\sum_{i \in \mathbf{n}}p_i \geq \sum_{i \in \mathbf{n}}\sum_{j \in \mathcal{A}_{i}}u_{ji}.$$ Hence contradiction. Therefore, given an $U \in \mathcal{U}$, there must exist $i \in \mathbf{n}$, $x_i(U) = \{\text{safe}, \text{precarious}\}$. In other words, in any power allocation matrix assuming any environment, it can never be the case that there is no survivor. 

$\qed$

Based on Lemma 1, an upper bound for the price of anarchy in the PAG is immediate. In addition, environments which gives an lower bound for the price of anarchy can be constructed.

\begin{theorem} In $\Gamma$,
$$1 \leq \text{PoA} \leq \frac{A}{B}$$ where $$A	= n \text{sup}\{\sum_{j \in \mathcal{F}_i}t_{ij}^{\mathcal{F}}(1) + \sum_{j \in \mathcal{A}_i}t_{ij}^{\mathcal{A}}(1): i \in \mathbf{n}\}$$ and $$B = (n-1)\text{inf}\{t_{ii}^{\mathcal{F}}(0): i \in \mathbf{n}\} + \text{inf}\{t_{ii}^{\mathcal{F}}(1): i \in \mathbf{n}\}.$$

\end{theorem}

\noindent {\it Proof of Theorem 4:} In an environment without any antagonism among countries, 
$\displaystyle\max_{U \in \mathcal{U}} \sum_{i \in \mathbf{n}}f^*_{i}(U)$ is achieved with all countries allocating zero to one other. At the same time, $\displaystyle\max_{U^* \in \mathcal{U^*}}\sum_{i \in \mathbf{n}}f^*_{i}(U^*)$ is also achieved because this is obviously an equilibrium. Therefore, in this case $\text{PoA} = 1$. 

Obviously, there exists no lower value for $\text{PoA}$; otherwise, it means that even better total welfare can be achieved in equilibrium. However, this gives a contradiction because $U^*$ already achieves the optimal total welfare. 1 is tight as a lower bound for the PAG in environments without any adversary relations. 

Now rank all the pairwise utility from not having survived itself of all countries, $t_{ii}(0)$, $i \in \mathbf{n}$, nondecreasingly, and denote the maximum as $$\text{sup}\{t_{ii}^{\mathcal{F}}(0): i \in \mathbf{n}\}$$ and the minimum as $$\text{inf}\{t_{ii}^{\mathcal{F}}(0): i \in \mathbf{n}\}.$$

Rank all the pairwise utilities from having survived itself of all countries, $t_{ii}^{\mathcal{F}}(1), i \in \mathbf{n}$, nondecreasingly, and denote the maximum as $$\text{sup}\{t_{ii}^{\mathcal{F}}(1): i \in \mathbf{n}\}$$ and the minimum as $$\text{inf}\{t_{ii}^{\mathcal{F}}(1): i \in \mathbf{n}\}.$$

Rank the optimal welfare of all countries $$\sum_{j \in \mathcal{F}_{i}}t_{ij}^{\mathcal{F}}(1) + \sum_{j \in \mathcal{A}_i}t_{ij}^{\mathcal{A}}(1), i \in \mathbf{n}$$ nondecreasingly, and denote the highest as $$\text{sup}\{\sum_{j \in \mathcal{F}_i}t_{ij}^{\mathcal{F}}(1) + \sum_{j \in \mathcal{A}_i}t_{ij}^{\mathcal{A}}(1), i \in \mathbf{n}\}$$. 
By Lemma 1, in any PAG, at least one country survives. Suppose there exists a PAG where exactly one country survives and the utility of this country is $\text{inf}\{t_{ii}^{\mathcal{F}}(1): i \in \mathbf{n}\}$. Since the other countries have not survived, their utilities are at least $$\text{inf}\{t_{ii}^{\mathcal{F}}(0): i \in \mathbf{n}\}.$$ 

This then gives the upper bound $\text{PoA} \leq \frac{A}{B}$ where $$A	= n \text{sup}\{\sum_{j \in \mathcal{F}_i}t_{ij}^{\mathcal{F}}(1) + \sum_{j \in \mathcal{A}_i}t_{ij}^{\mathcal{A}}(1): i \in \mathbf{n}\}$$ and $$B = (n-1)\text{inf}\{t_{ii}^{\mathcal{F}}(0): i \in \mathbf{n}\} + \text{inf}\{t_{ii}^{\mathcal{F}}(1): i \in \mathbf{n}\}.$$

$\qed$

% The upper bound is tight for the power allocation game assuming \emph{certain} networked international environment, such as the one in Example 2. 

% \begin{tikzpicture}[x=4em,y=-4em][scale=0.25]
% 	\drawnodex{-1,-1/2}{v1}{left}{v1}{above}{$8$}
% 	\drawnodex{-1,1/2}{v2}{left}{v2}{below}{$6$}
% 	\drawnodex{0,0}{v3}{below}{v3}{above}{$4$}
% 	\drawnodex{1,0}{v4}{below}{v4}{above}{$2$}
			
%     \drawfoe{v1}{v2}%{$w_{1,3}$}{$w_{3,1}$}
% 	\drawfoe{v2}{v3}%{$w_{2,3}$}{$w_{3,2}$}
% 	\drawfoe{v3}{v4}%{$w_{1,2}$}{$w_{2,1}$}

% \end{tikzpicture}
% \qquad
% \begin{tikzpicture}[x=4em,y=-4em][scale=0.25]
% 	\drawnodex{-1,-1/2}{v1}{left}{v1}{above}{$0$}
% 	\drawnodex{-1,1/2}{v2}{left}{v2}{below}{$0$}
% 	\drawnodex{0,0}{v3}{below}{v3}{above}{$0$}
% 	\drawnodex{1,0}{v4}{below}{v4}{above}{$2$}
			
%     \drawfoex{v1}{v2}{$8$}{$0$}
% 	\drawfoex{v2}{v3}{$6$}{$0$}
% 	\drawfoex{v3}{v4}{$4$}{$0$}

% \end{tikzpicture}

% \begin{example}

% \end{example}

%\section{Real World Examples}

\section{Discussion and Conclusion}

This paper analytically studies a paradox emerging from the PAG. Specifically, the paper shows friends may play different roles in a country's survival and its attainment of optimal welfare. Much like what Example 1 has shown, a country's having many friends may impede the attainment of its optimal welfare from power allocation, especially the potential friends have conflicts among themselves. 

However, paradoxes of this kind is unsurprising in a political context. In order to win over as many allies as possible always requires a country to straddle middle grounds between parties with perhaps irreconcilable differences or even conflicts. Just as the former British PM, Margaret Thatcher, accurately put it, ``standing in the middle of the road is very dangerous; you get knocked down by the traffic from both sides.'' Especially, thinking of the current conflictual scenarios between the United States and North Korea, a question can perhaps be asked, should China try to reconcile both or choose to exert pressure on one of them, e.g., North Korea?

\bibliography{paradox}
\bibliographystyle{unsrt}

\end{document}